\documentclass[12pt]{iopart}

\usepackage{graphicx}
\usepackage{color}
\usepackage{dcolumn}

\begin{document}

\title{Resonances for symmetric two--barrier potentials}
\author{Francisco M. Fern\'{a}ndez}

\address{INIFTA (UNLP, CCT La Plata-CONICET), Divisi\'on Qu\'imica Te\'orica,
Blvd. 113 S/N,  Sucursal 4, Casilla de Correo 16, 1900 La Plata,
Argentina}\ead{fernande@quimica.unlp.edu.ar}

\maketitle

\begin{abstract}
We describe a method for the accurate calculation of bound--state
and resonance energies for one--dimensional potentials. We
calculate the shape resonances for symmetric two--barrier
potentials and compare them with those coming from the Siegert
approximation, the complex scaling method and the
box--stabilization method. A comparison of the Breit--Wigner
profile and the transmission coefficient about its maximum
illustrates that the agreement is better the sharper the
resonance.
\end{abstract}

\section{Introduction}

\label{sec:intro}

In a recent paper Rapedius\cite{R11} showed how to calculate resonance
positions and widths by means of the Siegert approximation (SA). He applied
it to two exactly solvable models and also to the transmission through a
double barrier. In the latter nontrivial example he compared the approximate
SA resonances with the more accurate ones provided by complex scaling (CS).
In this interesting pedagogical paper Rapedius showed that the SA is
suitable for narrow resonances. He also described the difference between
Siegert and transmission resonances. The former are complex eigenvalues of
the Schr\"{o}dinger equation whereas the latter are related to the maxima of
the transmission coefficient.

Somewhat earlier Dutt and Kar\cite{DK10} had discussed scattering through
smooth double barriers constructed by means of Gaussian functions. They
compared the accurate transmission coefficient calculated numerically with
analytical expressions derived by means of the WKB method. The agreement is
remarkable for all values of the energy of the incident particle.

The physical interest in double--barrier potentials arises from models for
fission barrier\cite{WB69,CN70} as well as in the study of chemical reaction
thresholds\cite{FT91} and simple molecular collisions\cite{RL93}. In such
cases sharp resonances are associated to states trapped between the two
barriers\cite{CN70,FT91} and exhibit Lorentzian or Breit--Wigner (BW) profile%
\cite{RL93}. In addition to its remarkably accurate results\cite{CN70} the
WKB method is suitable for proving that the resonances are of BW type\cite
{C68,M98}. Square wells and barriers have also proved suitable for a
pedagogical discussion of bound states, virtual states and resonances\cite
{WHZ82}.

The purpose of this paper is to illustrate the relation between Siegert and
transmission resonances in a somewhat more detailed way by means of the
symmetrical two--barrier potential discussed by Rapedius\cite{R11}. The
Schr\"{o}dinger equation for this potential is not exactly solvable but one
can calculate the Siegert resonances accurately by means of the
Riccati--Pad\'{e} method (RPM)\cite{F95,F96}. It is also possible to obtain
the transmission coefficient as a function of the energy by means of the
Wronskian method\cite{F11b,F11c} and thus to compare its resonance peaks
with the BW expression for different barrier heights. This paper is expected
to be a useful complement to those earlier pedagogical discussions on the
subject\cite{R11,DK10,WHZ82} and a suitable approach to the problem for
courses of quantum mechanics at advanced undergraduate or graduate level.

In section \ref{sec:Siegert} we derive the SA in a way that differs from
that followed by Rapedius\cite{R11}. Like this author we introduce the
Siegert and transmission resonances that will be discussed and compared in
later sections. In section \ref{sec:RPM} we outline the main ideas of the
RPM for symmetric one--dimensional potentials. In section \ref{sec:results}
we calculate the Siegert resonances by means of the RPM and the transmission
coefficient by means of the Wronskian method. We can thus compare the peak
of the transmission coefficient with the BW profile for several barrier
heights. Finally, in section \ref{sec:conclusions} we summarize the main
results and draw conclusions.

\section{The Siegert approximation method}

\label{sec:Siegert}

Rapedius\cite{R11} developed the SA from the time--dependent Schr\"{o}dinger
equation. Since this equation is not used in the calculation of neither the
Siegert nor the transmission resonances we think that it may be fruitful to
derive the SA entirely from the time--independent Schr\"{o}dinger equation.

In order to simplify and facilitate both the algebra and the numerical
calculations in this paper we first convert the Schr\"{o}dinger equation
into a dimensionless eigenvalue equation. In this way one removes all the
physical constants and reduces the number of model parameters to a minimum.
Although we have already discussed this well known procedure in earlier
papers\cite{F11b,F11c,F11a} we think that it is worthwhile to insist on the
advantages of working with proper dimensionless equations in physics.

The time--independent Schr\"{o}dinger equation for a particle of mass $m$
that moves in one dimension ($-\infty <X<\infty $) under the effect of a
potential $V(X)$ is
\begin{equation}
-\frac{\hbar ^{2}}{2m}\psi ^{\prime \prime }(X)+V(X)\psi (X)=E\psi (X)
\label{eq:Schrodinger}
\end{equation}
where a prime indicates derivative with respect to the coordinate $X$. If we
define the dimensionless coordinate $x=X/L$, where $L$ is an appropriate
length scale (or length unit), then we obtain the dimensionless eigenvalue
equation
\begin{eqnarray}
&&-\frac{1}{2}\varphi ^{\prime \prime }(x)+v(x)\varphi (x)=\epsilon \varphi
(x)  \nonumber \\
&&\varphi (x)=\sqrt{L}\psi (Lx),\;v(x)=\frac{mL^{2}}{\hbar ^{2}}%
V(Lx),\;\epsilon =\frac{mL^{2}E}{\hbar ^{2}}  \label{eq:Schro_dim}
\end{eqnarray}
The length unit $L$ that renders both $\epsilon $ and $v(x)$ dimensionless
is arbitrary and we can choose it in such a way that makes the
Schr\"{o}dinger equation as simple as possible as shown in the first example
of section \ref{sec:results}.

For the time being we just assume that the potential tends to zero
\begin{equation}
\lim_{|x|\rightarrow \infty }v(x)=0
\end{equation}
faster than $|x|^{-1}$ and that the particle approaches the interaction
region from the left. Therefore, the boundary conditions are
\begin{eqnarray}
\lim_{x\rightarrow -\infty }\varphi (x) &=&\varphi _{-}(x)=Ae^{ikx}+Be^{-ikx}
\nonumber \\
\lim_{x\rightarrow \infty }\varphi (x) &=&\varphi _{+}(x)=Ce^{ikx}
\label{eq:scat_bc}
\end{eqnarray}
where $k=\sqrt{2\epsilon }$.

It follows from the Schr\"{o}dinger equation (\ref{eq:Schro_dim}) and its
complex conjugate that
\begin{equation}
\left( \varphi ^{\prime }\varphi ^{*}-\varphi ^{\prime *}\varphi \right)
^{\prime }=2\left( \epsilon ^{*}-\epsilon \right) |\varphi |^{2}
\label{eq:W'}
\end{equation}
If $\epsilon $ is real then the Wronskian $W(\varphi ^{*},\varphi )=\varphi
^{\prime }\varphi ^{*}-\varphi ^{\prime *}\varphi $ is constant and the
substitution of the asymptotic solutions (\ref{eq:scat_bc}) into $W(\varphi
_{-}^{*},\varphi _{-})=W(\varphi _{+}^{*},\varphi _{+})$ leads to
\begin{equation}
|A(\epsilon )|^{2}-|B(\epsilon )|^{2}=|C(\epsilon )|^{2}
\end{equation}
or $T(\epsilon )+R(\epsilon )=1$ where $T(\epsilon )=|C(\epsilon
)|^{2}/|A(\epsilon )|^{2}$ and $R(\epsilon )=|B(\epsilon )|^{2}/|A(\epsilon
)|^{2}$ are the transmission and reflection coefficients, respectively.

Following Rapedius\cite{R11} we call transmission resonance to an energy
value $\epsilon =\epsilon _{T}$ for which $T(\epsilon )$ exhibits a local
maximum. In the particular case that the potential is symmetric about the
origin $v(-x)=v(x)$ the maximum has unit value $T(\epsilon _{T})=1$\cite{C68}%
; consequently, $R(\epsilon _{T})=0$, $|A(\epsilon _{T})|^{2}=|C(\epsilon
_{T})|^{2}$ and $|B(\epsilon _{T})|^{2}=0$. The boundary conditions for such
particular energy value are
\begin{equation}
\varphi _{T\pm }^{\prime }(x)=ik_{T}\varphi _{T\pm }(x)
\end{equation}
and
\begin{equation}
|\varphi _{T-}|^{2}=|A_{T}|^{2}=|C_{T}|^{2}=|\varphi _{T+}|^{2}
\end{equation}
where the subscript $T$ indicates that $\epsilon =\epsilon _{T}$. For
concreteness we assume that the potential is symmetric from now on. If it
were symmetric about another point $x=x_{0}\neq 0$ we would simply shift the
coordinate origin from $x=0$ to $x=x_{0}$.

A resonance (or Siegert) eigenstate is a solution to the Schr\"{o}dinger
equation that behaves asymptotically as a purely outgoing wave\cite{S39,C68}
(and references therein):

\begin{eqnarray}
\lim_{x\rightarrow -\infty }\varphi _{S}(x) &=&\varphi
_{S-}(x)=B_{S}e^{-ik_{S}x}  \nonumber \\
\lim_{x\rightarrow \infty }\varphi _{S}(x) &=&\varphi
_{S+}(x)=C_{S}e^{ik_{S}x}  \label{eq:Siegert_bc}
\end{eqnarray}
where the subscript $S$ indicates that $\epsilon =\epsilon _{res}$
which is a complex eigenvalue $\epsilon _{res}=\epsilon
_{R}+i\epsilon _{I}$. The real part is the resonance position and
the imaginary part is related to the resonance width $\Gamma
=-2\epsilon _{I}>0$. For practical purposes it is customary to
assume that these boundary conditions are approximately valid for
a sufficiently large coordinate value $|x|=a$. Therefore, if we
integrate equation (\ref {eq:W'}) between $-a$ and $a$ we obtain
\begin{eqnarray}
\left. \left( \varphi ^{\prime }\varphi ^{*}-\varphi ^{\prime *}\varphi
\right) \right| _{-a}^{a} &=&i(k_{S}+k_{S}^{*})\left( |\varphi
_{S+}|^{2}+|\varphi _{S-}|^{2}\right)  \nonumber \\
&=&-4i\epsilon _{I}\int_{-a}^{a}|\varphi _{S}|^{2}\,dx
\end{eqnarray}
The Siegert states for a symmetrical potential are either even or odd;
therefore $|\varphi _{S+}|^{2}=|\varphi _{S-}|^{2}$. In the case of a
sufficiently narrow resonance $|\epsilon _{I}|\ll \epsilon _{R}$ we may
safely carry out the additional approximation that $\epsilon _{res}\approx
\epsilon _{R}\approx \epsilon _{T}$.

Since the Siegert state is strongly localized in the well between the
barriers located at $x=\pm b$\cite{R11} we can also write
\begin{equation}
\int_{-a}^{a}|\varphi _{S}|^{2}\,dx\approx \int_{-b}^{b}|\varphi
_{S}|^{2}\,dx\approx \int_{-b}^{b}|\varphi _{T}|^{2}\,dx
\end{equation}
Finally, from the equations above we obtain an expression for the resonance
width:
\begin{equation}
\Gamma =\frac{k_{T}|\varphi _{T}(a)|^{2}}{\int_{0}^{b}|\varphi _{T}|^{2}\,dx}
\label{eq:SA}
\end{equation}
already derived by Rapedius\cite{R11} by means of the time--dependent
Schr\"{o}dinger equation.

A more rigorous, general and elegant derivation of the results above was
given by Whitton and Connor\cite{WC73} many years ago by means of a
Wronskian analysis. Those authors called the $S$ and $T$ boundary conditions
``outward moving waves only'' and ``forward moving waves only'',
respectively. Rapedius\cite{R11} also showed an expression for the
nonsymmetric case $v(-x)\neq v(x)$ that we do not consider here. The general
expressions of Whitton and Connor\cite{WC73} do in fact apply to both the
symmetric and nonsymmetric case. Note that the energies for the
forward--moving--waves--only boundary conditions are real for a symmetric
potential but they may be complex for a nonsymmetric one \cite{WC73}.

It is clear from all the assumptions made above that the SA equation (\ref
{eq:SA}) applies only to sufficiently narrow resonances. In fact, Rapedius
\cite{R11} illustrated this point by comparing SA results with those
provided by the more accurate CS. In section~\ref{sec:results} we propose an
alternative comparative discussion of Siegert and transmission resonances.

\section{The Riccati--Pad\'{e} method (RPM)}

\label{sec:RPM}

In order to compare Siegert and transmission resonances we need sufficiently
accurate complex eigenvalues of the Schr\"{o}dinger equation with purely
outgoing--wave boundary conditions. There are many suitable methods for this
purpose; for example, Rapedius\cite{R11} discussed the SA and CS ones. In
what follows we outline the RPM that yields remarkably accurate results for
narrow and broad resonances\cite{F95,F96} (a pedagogical approach to RPM for
bound states is available elsewhere\cite{F08}). We think that the RPM may be
a suitable practical tool for a quantum mechanics course because the
derivation of its main equations and their implementation in a computer
program are both straightforward.

Suppose that we want to obtain the eigenvalues of the dimensionless
Schr\"{o}dinger equation (\ref{eq:Schro_dim}) with a symmetric potential $%
v(-x)=v(x)$ (without loss of generality we assume that $v(0)=0$). We
restrict ourselves to those eigenstates that are either even $\varphi
(-x)=\varphi (x)$ or odd $\varphi (-x)=-\varphi (x)$ (both bound and
resonance states satisfy this criterion). We define the regularized
logarithmic derivative of the eigenfunction
\begin{equation}
f(x)=\frac{s}{x}-\frac{\varphi ^{\prime }(x)}{\varphi (x)}
\label{eq:log_der}
\end{equation}
where $s=0$ or $s=1$ for even or odd states, respectively. It satisfies the
Riccati equation
\begin{equation}
f^{\prime }(x)+\frac{2s}{x}f(x)-f(x)^{2}+2v(x)-2\epsilon =0
\label{eq:Riccati}
\end{equation}

If we can expand $v(x)$ in a Taylor series about $x=0$%
\begin{equation}
v(x)=\sum_{j=1}^{\infty }v_{j}x^{2j}  \label{eq:V_series}
\end{equation}
then we can also expand $f(x)$ about the same point as
\begin{equation}
f(x)=x\sum_{j=0}^{\infty }f_{j}(\epsilon )z^{j},\;z=x^{2}
\label{eq:f(x)_series}
\end{equation}
Note that the term $s/x$ in equation (\ref{eq:log_der}) removes the pole at
origin in the case of odd states ($\varphi ^{odd}(0)=0$). The expansion of
equation (\ref{eq:Riccati}) in a Taylor series about the origin leads to a
recurrence relation for the coefficients $f_{j}$ that enables us to obtain
as many coefficients $f_{j}(\epsilon )$ as necessary:
\begin{eqnarray}
f_{n} &=&\frac{1}{2n+2s+1}\left( \sum_{j=0}^{n-1}f_{j}f_{n-j-1}+2\epsilon
\delta _{n0}-2v_{n}\right) ,\;n=1,2,\ldots  \nonumber \\
f_{0} &=&\frac{2\epsilon }{2s+1}  \label{eq:fn}
\end{eqnarray}

Since $f(x)$ has poles at the zeros of $\varphi (x)$ we look for a rational
approximation of the form $f(x)\approx x[M/N](z)$, where
\begin{equation}
\lbrack M/N](z)=\frac{\sum_{j=0}^{M}a_{j}z^{j}}{\sum_{j=0}^{N}b_{j}z^{j}}%
=\sum_{j=0}^{M+N+1}f_{j}(\epsilon )z^{j}+O(z^{M+N+2})  \label{eq:[M/N](z)}
\end{equation}
Because we can arbitrarily choose $b_{0}=1$ we are left with $M+N+1$
coefficients $a_{j}$ and $b_{j}$ of the rational function and the unknown
energy as independently adjustable parameters. Therefore we require that the
rational approximation (Pad\'{e} approximant) yields $M+N+2$ exact
coefficients of the Taylor series for $f(x)$ as explicitly indicated in
equation (\ref{eq:[M/N](z)}). When $M\geq N$ we easily derive the following
equations:
\begin{eqnarray}
\sum_{k=0}^{\min (j,N)}b_{k}f_{j-k} &=&a_{j},\;j=0,1,\ldots ,M  \nonumber \\
\sum_{k=0}^{N}b_{k}f_{j-k} &=&0,\;j=M+1,M+2,\ldots ,M+N+1  \label{eq:aj_bj}
\end{eqnarray}
We can view the second set of equations as a system of $N+1$ homogeneous
equations with $N+1$ unknowns $b_{N},b_{N-1},\ldots ,b_{0}$. Therefore,
there will be a nontrivial solution only if $\epsilon $ is a root of
\begin{equation}
H_{D}^{d}(\epsilon )=\left|
\begin{array}{cccc}
f_{M-N+1} & f_{M-N+2} & \cdots & f_{M+1} \\
f_{M-N+2} & f_{M-N+3} & \cdots & f_{M+2} \\
\vdots & \vdots & \ddots & \vdots \\
f_{M+1} & f_{M+2} & \cdots & f_{M+N+1}
\end{array}
\right| =0  \label{eq:Hankel}
\end{equation}
where $d=M-N=0,1,\ldots $ and $D=N+1=2,3,\ldots $ is the dimension of the
Hankel determinant $H_{D}^{d}(\epsilon )$. As $D$ increases, sequences of
roots $\epsilon ^{[D,d]}$ of the Hankel determinant converge toward the
allowed energies of the Schr\"{o}dinger equation. This sort of quantization
condition applies to bound states and resonances\cite{F95,F96,F08}. We
simply identify convergent sequences of real and complex roots $\epsilon
^{[D,d]}$ of the Hankel determinants $H_{D}^{d}(\epsilon )$ for $%
D=2,3,\ldots ,D_{m}$ and estimate the error of the calculation as $\left|
w^{[D_{m}]}-w^{[D_{m}-1]}\right| $, where $w^{[D]}$ stands for the real or
complex part of $\epsilon ^{[D,d]}$. We thus truncate the results to the
last stable digit.

\section{Results and discussion}

\label{sec:results}

As a first illustrative example we consider the symmetrical double barrier $%
V(X)=V_{0}X^{2}e^{-\alpha X^{2}}$, where $V_{0},\alpha >0$, discussed by
Rapedius\cite{R11}. The Schr\"{o}dinger equation for this potential is not
exactly solvable but it is sufficiently simple for pedagogical purposes. In
this case the dimensionless potential is given by
\begin{eqnarray}
v(x) &=&v_{0}x^{2}e^{-\lambda x^{2}},  \nonumber \\
v_{0} &=&\frac{mL^{4}V_{0}}{\hbar ^{2}},\;\lambda =\alpha L^{2}
\label{eq:v(x)_DWB}
\end{eqnarray}
Note that without loss of generality we can treat this problem as a
one--parameter model because we may have either $v_{0}=1$ when $L^{2}=\hbar /%
\sqrt{mV_{0}}$ or $\lambda =1$ when $L^{2}=1/\alpha $. However, we write $%
v(x)$ as a two--parameter potential following Rapedius\cite{R11}. This
potential exhibits a well centered at $x=0$ and two barriers of height $%
v_{b}=v_{0}/(e\lambda )$ symmetrically located at $x=\pm b$, $b=1/\sqrt{%
\lambda }$.

For comparison purposes we first consider the potential parameters $%
v_{0}=1/2 $ and $\lambda =0.1$ already chosen by Rapedius\cite{R11}. Table~%
\ref{tab:resonances1} shows the first six resonances obtained from
Hankel sequences with $D\leq 20$ and $d=0$. The first three of
them agree with those calculated by means of SA and CS\cite{R11}.
Present results are supposed to be accurate to the last digit and,
consequently, much more accurate than the SA and CS ones shown by
Rapedius\cite{R11}. However, it is worth mentioning that the CS
results can in principle be made as accurate as desired by
increasing the dimension of the basis set\cite{KG02}. As stated
above (and already shown in Table~\ref{tab:resonances1}) the RPM
is suitable for both sharp and broad resonances. However, the
accuracy of the RPM for a given determinant dimension $D$
decreases with the resonance ``quantum number'' $n$ because the
Hankel sequences for higher resonances appear at greater values of
$D$.

The RPM is extremely accurate but it is not as general as other approaches
(like, for example, CS) because it only applies to separable Schr\"{o}dinger
equations. However, one--dimensional and separable models are widely
discussed in most courses on quantum mechanics.

It is not difficult to calculate the transmission probability $T(\epsilon )$
for the potential (\ref{eq:v(x)_DWB}) by means of the Wronskian method\cite
{F11b,F11c} and thus compare the Siegert $\epsilon _{res}$ and transmission $%
\epsilon _{T}$ resonances discussed by Rapedius\cite{R11}. More precisely,
we compare the actual transmission coefficient $T(\epsilon )$ and the BW
shape
\begin{equation}
T(\epsilon )\approx \frac{\epsilon _{I}^{2}}{(\epsilon -\epsilon
_{R})^{2}+\epsilon _{I}^{2}}  \label{eq:T_Lorentz}
\end{equation}
in a neighbourhood of the maximum of $T(\epsilon )$ at $\epsilon =\epsilon
_{T}\approx \epsilon _{R}$. Connor\cite{C68} has already shown that $%
T(\epsilon _{T})=1$ for a symmetric potential and that equation (\ref
{eq:T_Lorentz}) is a reasonable approximation for isolated sharp resonances.

\begin{figure}[H]
\begin{minipage}{6cm}
\includegraphics[width=6cm]{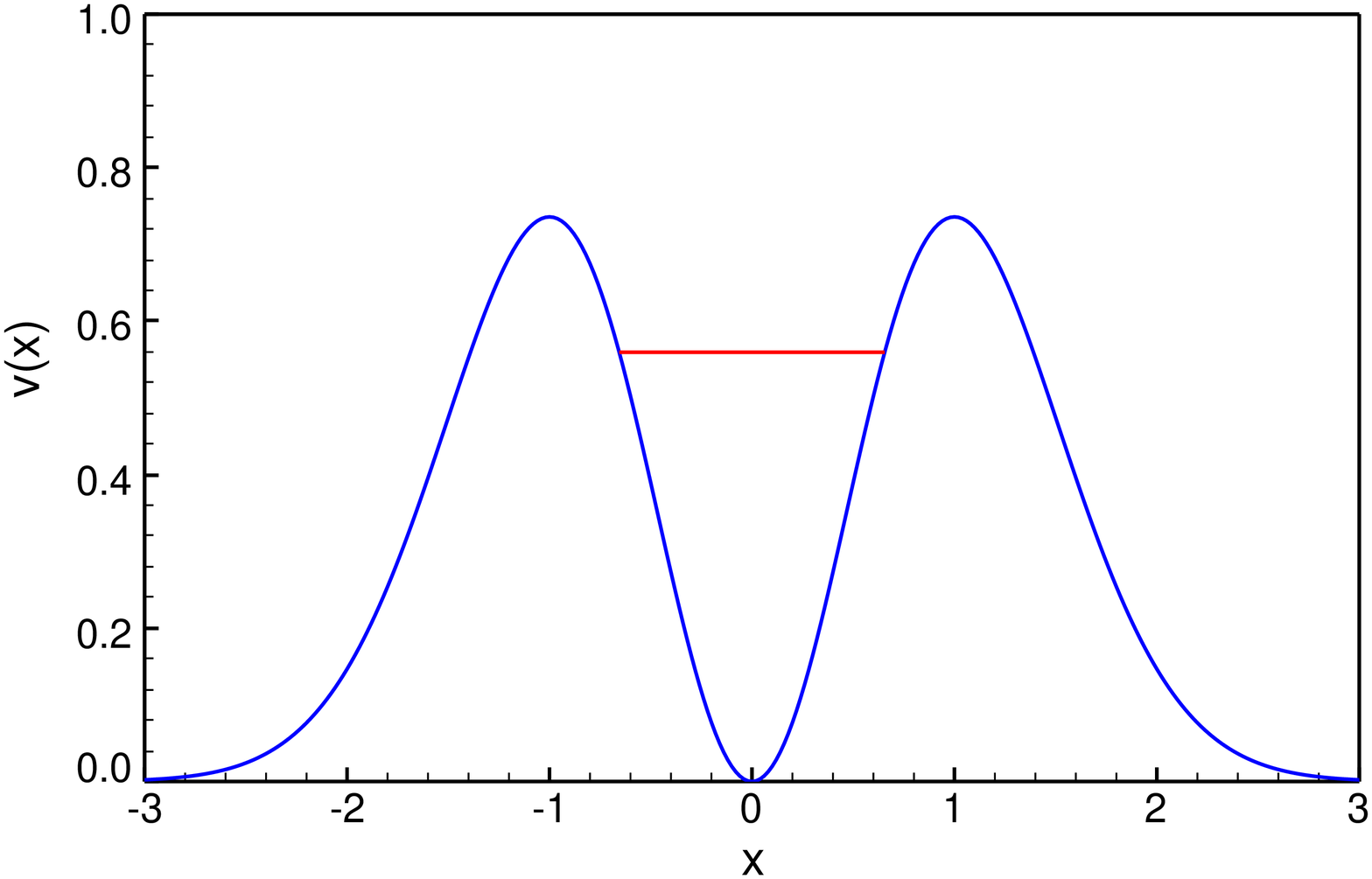}

\end{minipage}
\begin{minipage}{6cm}
\includegraphics[width=6cm]{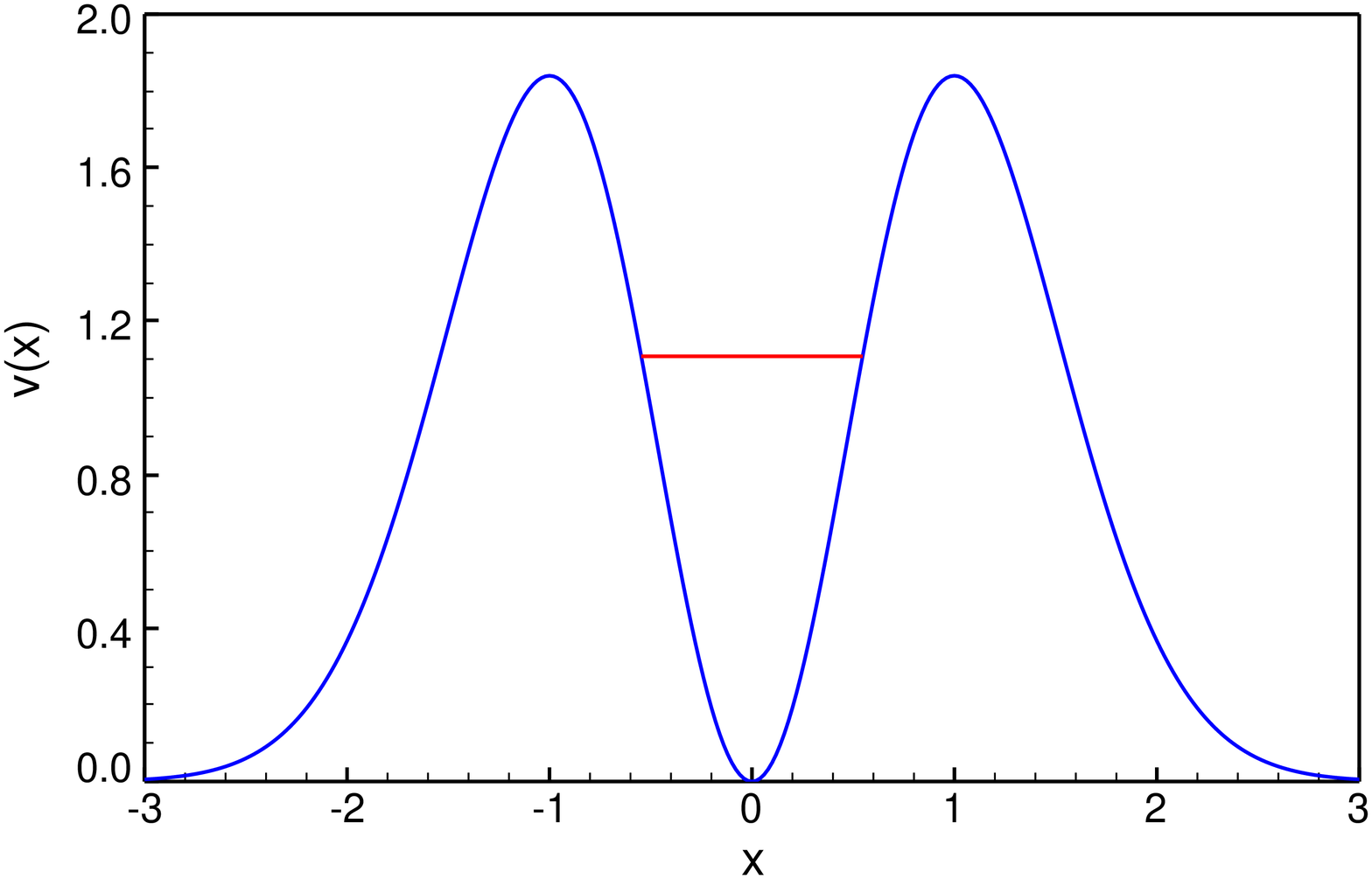}

\end{minipage}
\par
\begin{minipage}{6cm}
\includegraphics[width=6cm]{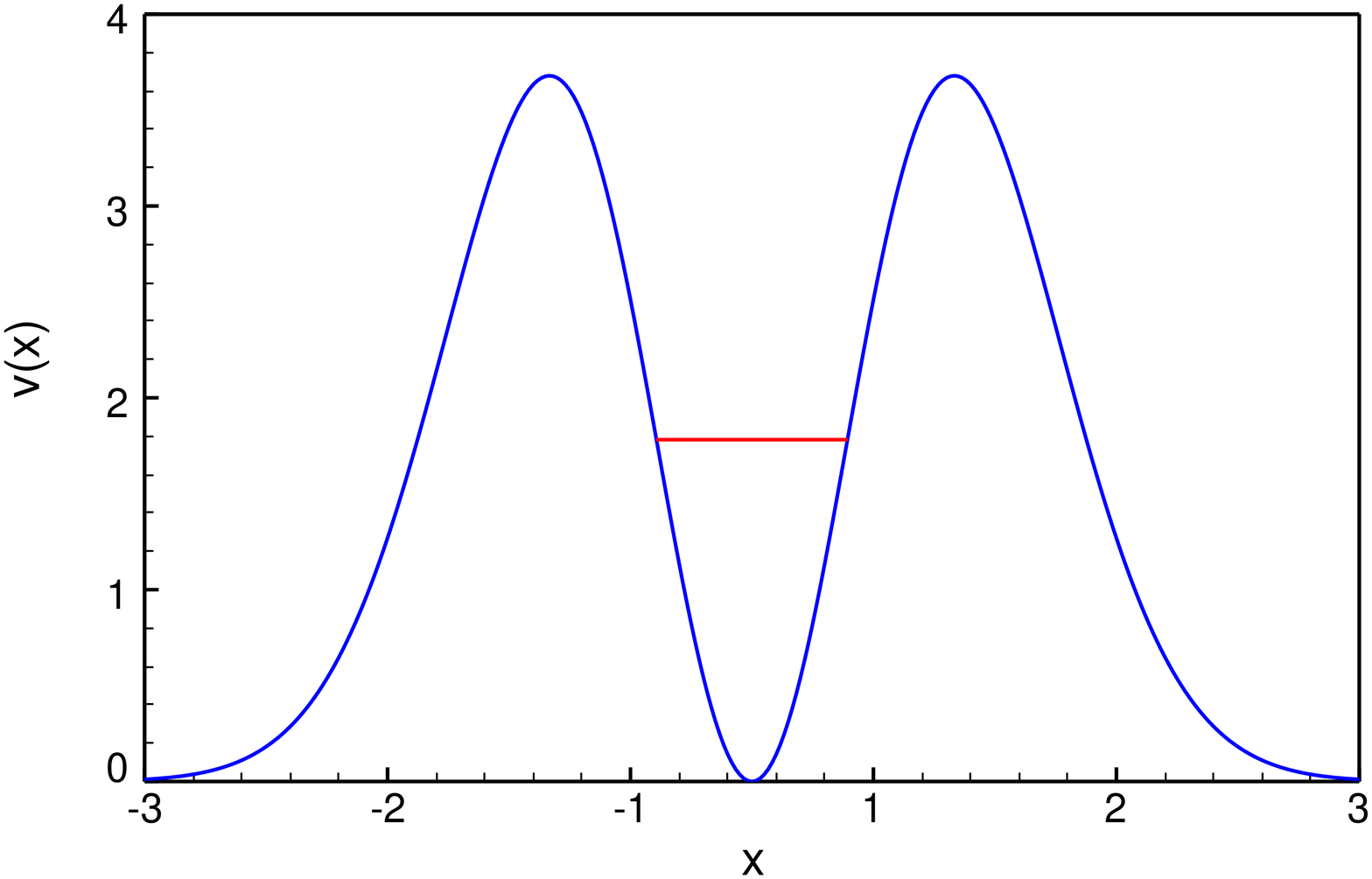}

\end{minipage}
\begin{minipage}{6cm}
\includegraphics[width=6cm]{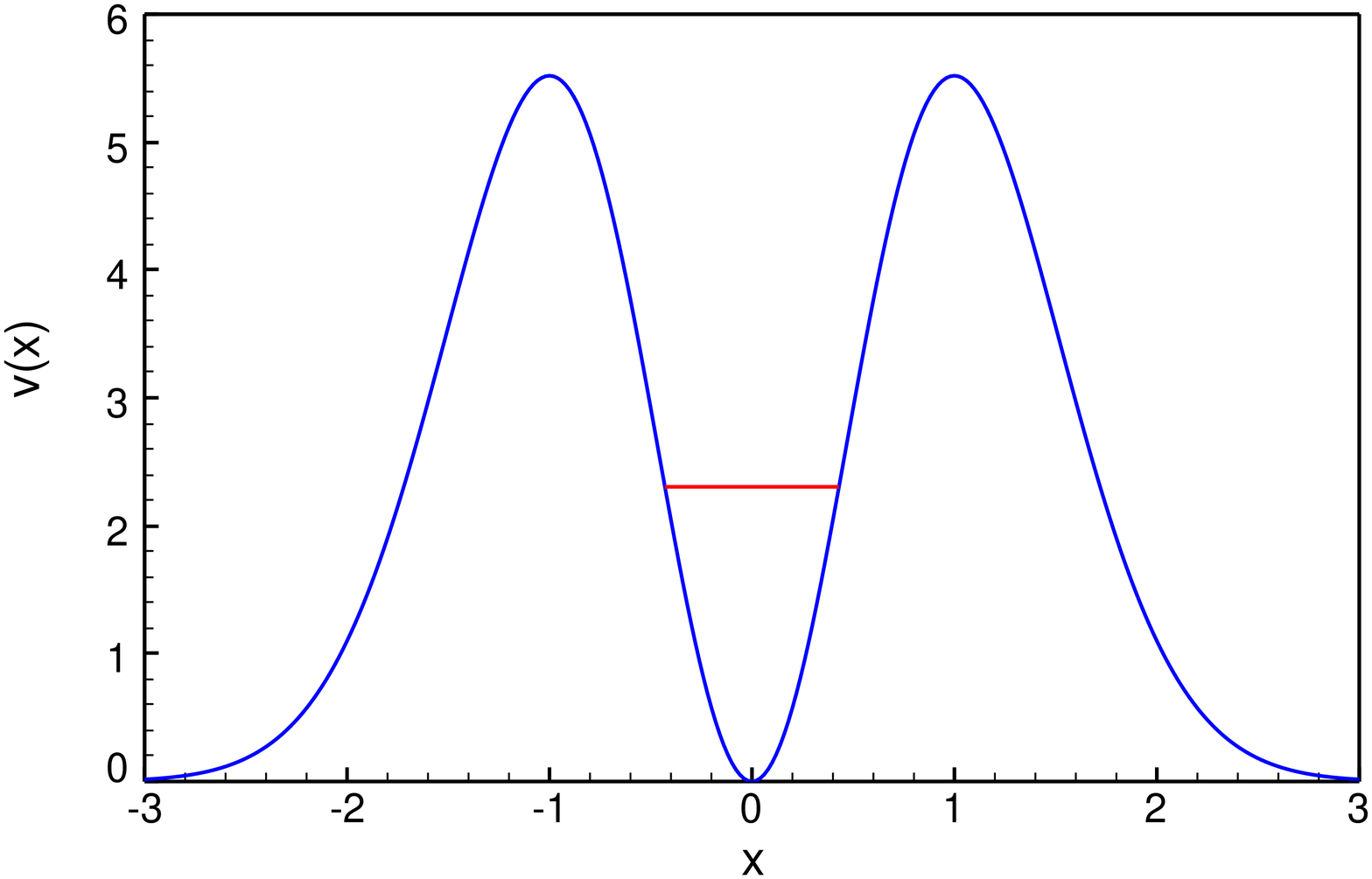}

\end{minipage}
\caption{(Color online) Potential function (blue) and quasi--stable energy
(red) for $v_0=2,\ 5,\ 10,\ 15$ }
\label{fig:v0_P}
\end{figure}

Table \ref{tab:resonance} shows the lowest resonance for $v_{0}=2,5,10,15$
and $\lambda =1$ calculated by means of the RPM exactly as discussed above.
We have rounded off all the results to the first unstable digit. The second
column displays the approximate values of the barrier heights $v_{b}=v_{0}/e$%
. Note that $|v_{b}-\epsilon _{R}|$ increases with $v_{0}$; that is to say,
the resonance moves deeper into the well between the barriers as $v_{0}$
increases. At the same time it becomes narrower and therefore more stable
(its decay rate $\Gamma =-2\epsilon _{I}$ decreases). Fig.~\ref{fig:v0_P}
shows the potential and the location of the lowest resonance (horizontal
line) for each case and illustrates graphically the behaviour just discussed.

\begin{figure}[H]
\begin{minipage}{6cm}
\includegraphics[width=6cm]{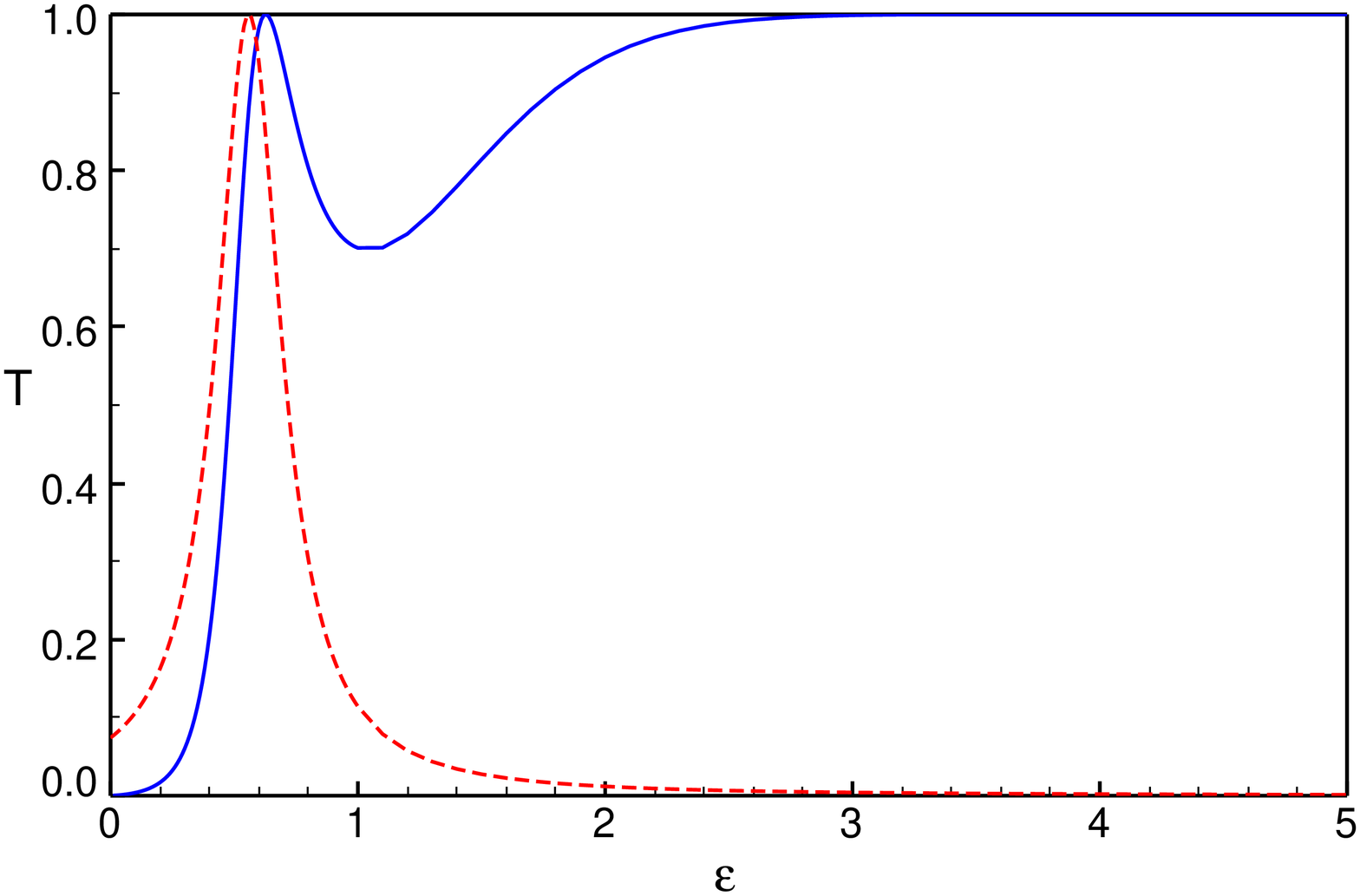}

\end{minipage}
\begin{minipage}{6cm}
\includegraphics[width=6cm]{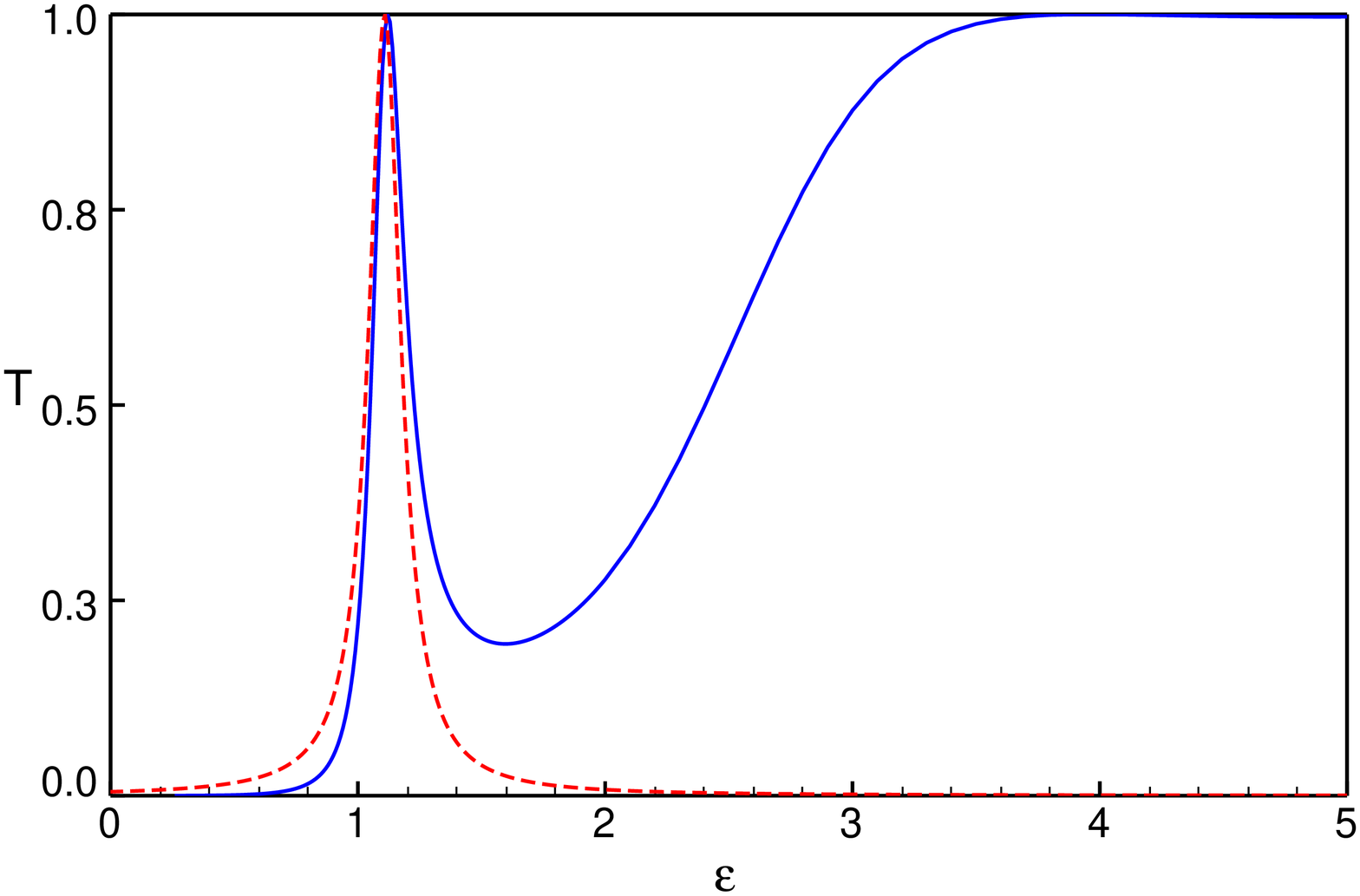}

\end{minipage}
\par
\begin{minipage}{6cm}
\includegraphics[width=6cm]{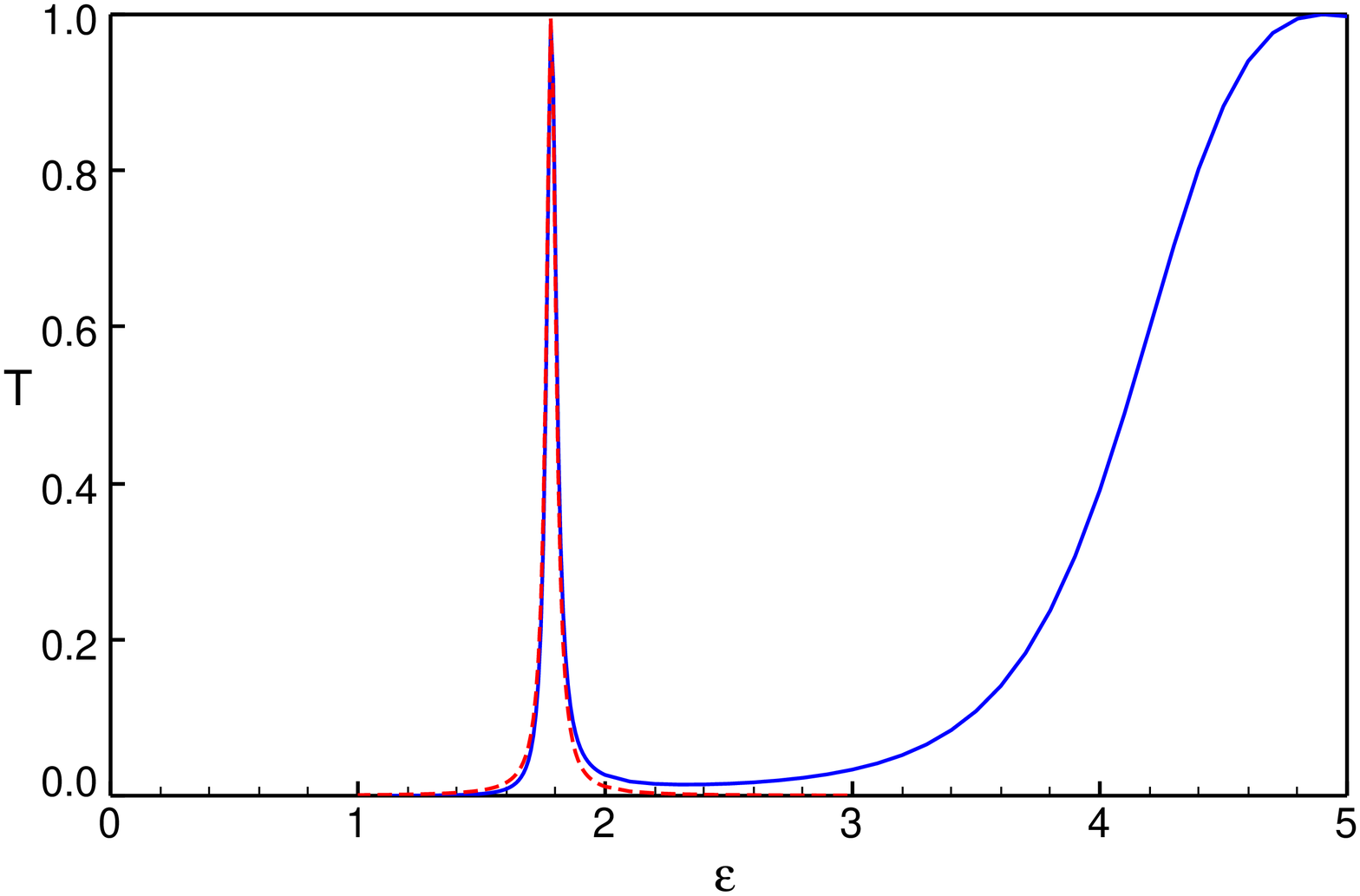}

\end{minipage}
\begin{minipage}{6cm}
\includegraphics[width=6cm]{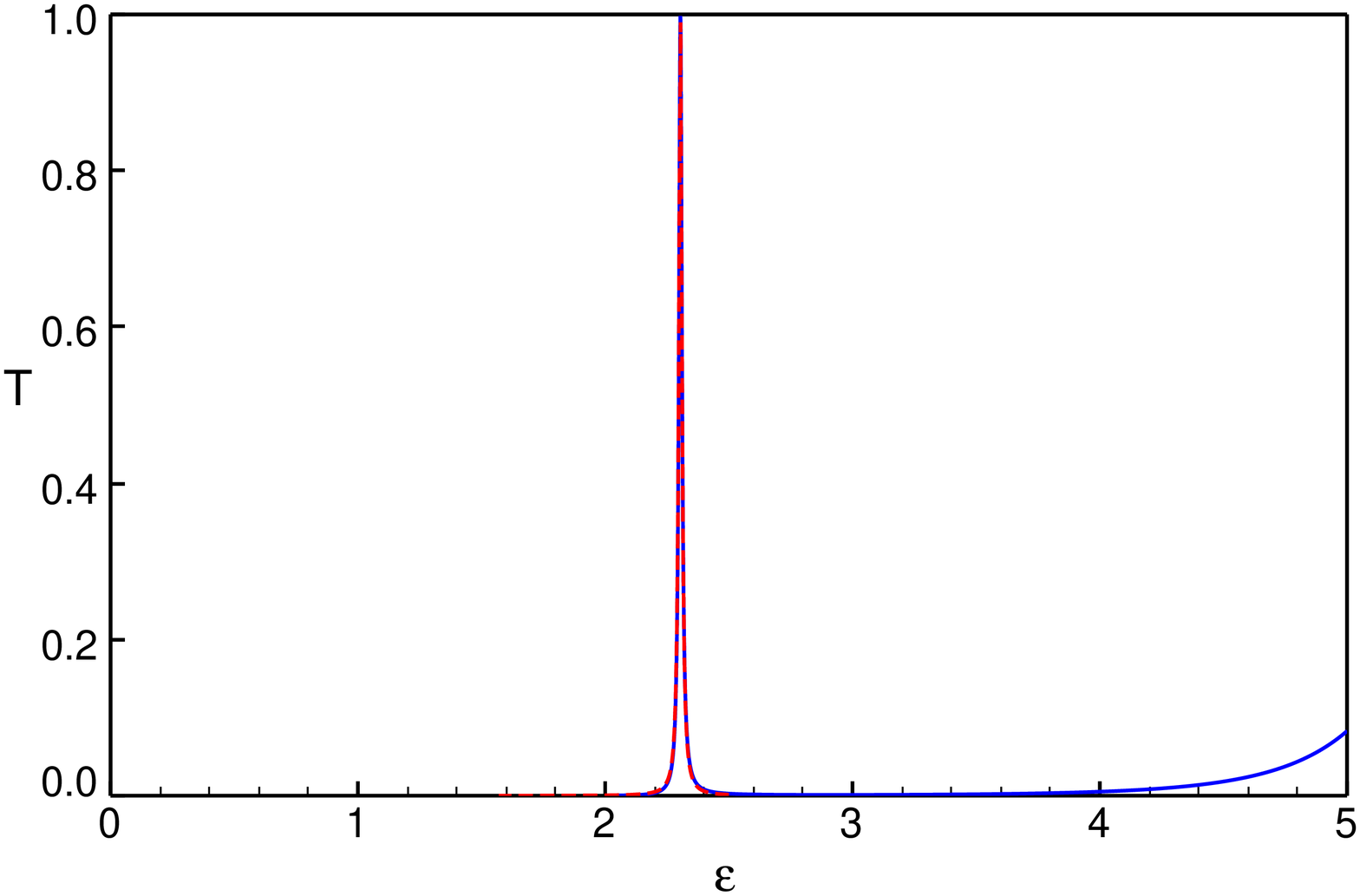}

\end{minipage}
\caption{(Color online) Numerical transmission probability (solid line,
blue) and Lorentzian profile (dashed line, red) for $v_0=2,\ 5,\ 10,\ 15$ }
\label{fig:T}
\end{figure}

Fig.~\ref{fig:T} shows the transmission probability $T(\epsilon )$
calculated by means of the Wronskian method\cite{F11b,F11c} and the BW
profile (\ref{eq:T_Lorentz}) with the values of $\epsilon _{R}$ and $%
\epsilon _{I}$ given in Table~\ref{tab:resonance}. It clearly illustrates
that $T(\epsilon )$ becomes sharper about the maximum as $\Gamma $
decreases. In addition to it, we appreciate the well known fact that the BW
profile gives a better description of the peak of the transmission
coefficient the sharper the resonance. In other words, as the potential
parameter increases from $v_{0}=2$ to $v_{0}=15$ and $\Gamma $ decreases the
BW profile (\ref{eq:T_Lorentz}) becomes increasingly more accurate in the
neighbourhood of the maximum.

\begin{figure}[H]
\begin{center}
\bigskip\bigskip\bigskip \includegraphics[width=9cm]{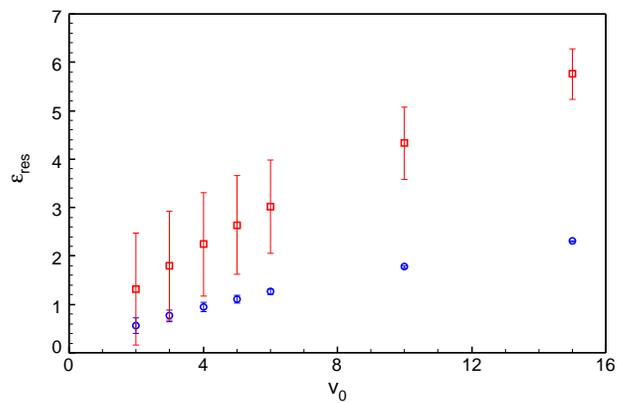}
\end{center}
\caption{(Color online) Real (symbols) and imaginary (error bars)
parts of the first (blue, circles) and second (red, squares)
resonance} \label{Fig:Res_ov}
\end{figure}

The BW profile is suitable for isolated resonances, and closely spaced broad
resonances may overlap. In such a case $T(\epsilon )$ becomes rather too
distorted for the BW profile to fit satisfactorily. Fig.~\ref{Fig:Res_ov}
shows the two first resonances for several values of $v_{0}$. The resonance
positions are indicated by symbols and their widths by error bars. We
clearly appreciate that the resonances overlap for $v_{0}=2$ and $v_{0}=3$
in which cases the BW profile is unsuitable. The magnitude of the overlap
diminishes with $v_{0}$ as the gap between the resonances increases and
their widths decrease.

Klaiman and Moiseyev\cite{KM10} have recently proposed an improved
profile based on the same information required for the BW one. The
KM profile is nonsymmetric and it is peaked at $|\epsilon
_{res}|>\epsilon _{R}$ which corrects the fact that typically
$\epsilon _{R}<\epsilon _{T}$ as shown in Fig.~\ref{fig:T}. We do
not show the KM profile here because the correction is mild for
the present model. The reason is that $\epsilon _{I}\ll \epsilon
_{R}$ and $ |\epsilon _{res}|\approx \epsilon _{R}$  for the
isolated resonances (say $v_{0}\geq 4$).

\begin{figure}[H]
\begin{center}
\bigskip\bigskip\bigskip \includegraphics[width=9cm]{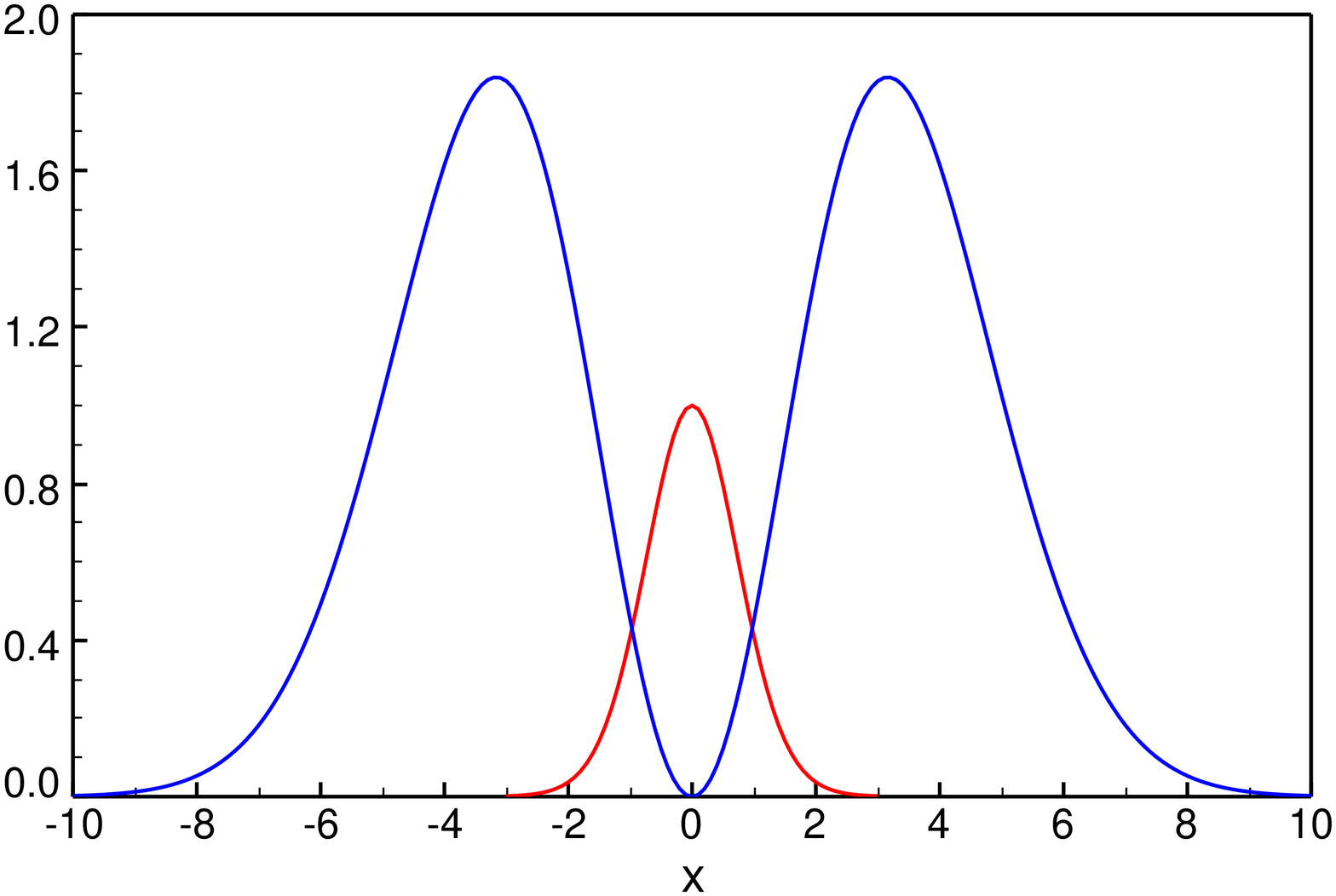}
\end{center}
\caption{(Color online) Potential (blue) and $|\varphi(x)|^2$ (red) for $%
v_0=1/2$ and $\lambda=0.1$}
\label{Fig:Phi}
\end{figure}

Once we have the complex energy eigenvalue we can easily calculate the
Siegert state by numerical integration or any other approach. For example,
its Taylor expansion about the origin
\begin{equation}
\varphi (x)=\sum_{j=0}^{\infty }c_{j}x^{2j+s}  \label{eq:phi_Taylor}
\end{equation}
provides a suitable approximate analytical expression. It is not difficult
to verify that the coefficients of this series are given by the recurrence
relation
\begin{equation}
c_{j+1}=\frac{2}{(2j+s+1)(2j+s+2)}\left[ \sum_{k=0}^{j}v_{k}c_{j-k}-\epsilon
c_{j}\right] ,\;j=0,1,\ldots
\end{equation}
where we arbitrarily choose $c_{0}=1$. The Taylor expansion (\ref
{eq:phi_Taylor}) converges for all values of $x$ but in practice we can only
add a finite number $M$ of terms and the resulting partial sum for $|\varphi
(x)|^{2}$ tends to infinity as $x\rightarrow \infty $. For this reason the
partial sum will be valid only in a finite interval $-x_{M}<x<x_{M}$ that we
should choose judiciously. Fig.~\ref{Fig:Phi} shows $v(x)$ for $v_{0}=1/2$
and $\lambda =0.1$ and $|\varphi (x)|^{2}$ for $M=24$. This figure is
similar to the one shown by Rapedius \cite{R11} for the same potential on a
wider abscissas interval. We clearly see that the Siegert state is so
strongly localized in the well between the two barriers that $|\varphi
(x)|^{2}\approx 0$ for $|x|>b$ which justifies one of the assumptions made
in section~\ref{sec:Siegert}.

Some time ago Korsch and Gl\"{u}ck\cite{KG02} proposed a pedagogical
approach to the calculation of the eigenvalues of the Schr\"{o}dinger
equation by means of the matrix representation of the coordinate and
momentum. Among other illustrative examples they considered the
one--dimensional potential
\begin{equation}
v(x)=\left( \frac{x^{2}}{2}-J\right) e^{-\lambda x^{2}}+J
\label{eq:v(x)_WDB2}
\end{equation}
which supports one bound state and many resonances when $J=0.8$ and $\lambda
=0.1$. Note that this potential--energy function reduces to a particular
case of Eq.~(\ref{eq:v(x)_DWB}) when $J=0$. Those authors calculated the
bound--state energy by a straightforward application of the matrix method
and the first two resonances by means of the box--stabilization method\cite
{KG02}. The application of the RPM to this problem is straightforward and
one simply looks for converging sequences of real and complex roots of the
Hankel determinants in order to obtain the energies of the bound--state and
resonances, respectively. Table~\ref{tab:resonances2} shows the eigenvalues
estimated from Hankel determinants of dimension $D\leq 20$. Also in this
case we see that the RPM results are remarkably accurate.

\section{Conclusions}

\label{sec:conclusions}

One of the purposes of this paper is to compare the Siegert and transmission
resonances in an alternative way to that discussed by Rapedius\cite{R11}. To
this end we calculated accurate Siegert resonances by means of the RPM and
the transmission coefficient by means of the Wronskian method\cite{F11b,F11c}
for the two--barrier potential (\ref{eq:v(x)_DWB}). The chosen model is
nontrivial but sufficiently simple for the straightforward aplication of
both approaches. In this way we can compare the shape of $T(\epsilon )$
about its maximum at $\epsilon =\epsilon _{T}$ (transmission resonance) with
the BW profile (\ref{eq:T_Lorentz}) constructed by means of the complex
eigenvalues $\epsilon _{res}=\epsilon _{R}+i\epsilon _{I}$ (Siegert
resonance). The comparison illustrates the well known fact that the BW
profile fits sharp isolated resonances more accurately. This conclusion
complements Rapedius observation that the SA becomes less accurate as the
resonance width increases\cite{R11}. In addition to it, we have argued that
in the present example the KM profile\cite{KM10} mildly improves on the BW
one because the lowest resonance is rather too sharp when it can be
considered sufficiently isolated from the next resonance.

We have derived the SA from the time--independent Schr\"{o}dinger equation
in a way that is more consistent with the methods for the calculation of the
resonances described in this and Rapedius' papers. The Wronskian approach is
most suitable for the discussion of the problem and enables the
decomposition of the resonance widths into the sum of partial widths\cite
{WC73} which we have omitted here.

We have shown that the RPM is a powerful tool for the location of resonances
in one--dimensional quantum--mechanical problems (more precisely, it applies
to separable Schr\"{o}dinger equations). The RPM is a local approach based
on the expansion of the logarithmic derivative of the wavefunction about a
chosen point, and for this reason it may not be the most convenient method
for a pedagogical illustration of the main features of the Siegert states
(for example, the discussion of the boundary conditions). However, it
provides remarkably accurate results and the derivation of its main
equations offers no difficulty. In addition to it, the recipe for its
application is simple and easy to encode into a computer program, especially
in most available computer algebra systems. We think that the RPM may be a
valuable tool for teaching purposes because it allows the students to obtain
accurate bound--state and resonance energies and compare them with those
provided by other approaches like SA, CS or the matrix method\cite{R11,KG02}.

\ack

I would like to thank Professor K. Rapedius for bringing Ref.~\cite{KM10} to
my attention and Professor S. Klaiman for useful suggestions.

\begin{table}[H]
\caption{Lowest resonances for the potential (\ref{eq:v(x)_DWB}) with $%
v_0=1/2$ and $\lambda=0.1$ }
\label{tab:resonances1}
\begin{center}
\begin{tabular}{D{.}{.}{3}D{.}{.}{20}D{.}{.}{20}}
\hline
  \multicolumn{1}{c}{$n$} & \multicolumn{1}{c}{$\epsilon_R$}
  & \multicolumn{1}{c}{$-\epsilon_I$}\\
\hline
0  &  0.46014727653933356360 & 9.6203883198201929683\times 10^{-7}       \\
1  &  1.2804203534682821470  & 1.6737132594145830404\times 10^{-3}      \\
2 &   1.8531086351750533910  & 6.7240255103872613345\times 10^{-2}       \\
3 &   2.2323252762455511600  & 0.33989855689185650713      \\
4 &   2.567615869399468602   & 0.8194028131702960163      \\
5 &   2.887957554267041665   & 1.409344599863779927          \\
\end{tabular}
\end{center}
\end{table}

\begin{table}[H]
\caption{Lowest resonance for the potential (\ref{eq:v(x)_DWB}) and several
potential parameters}
\label{tab:resonance}
\begin{center}
\begin{tabular}{D{.}{.}{2}D{.}{.}{3}D{.}{.}{20}D{.}{.}{20}}
\hline
  \multicolumn{1}{c}{$v_0$} & \multicolumn{1}{c}{$v_b$}&
  \multicolumn{1}{c}{$\epsilon_R$}
  & \multicolumn{1}{c}{$\epsilon_I$}\\
\hline
2  & 0.7358 &  0.55937118458252732995 &  0.15830525114271135525      \\
5  & 1.839  &  1.1082157629920295074  &  0.078972583905329832058     \\
10 & 3.679  &  1.7816763825869113601  &  0.023794309337967155927     \\
15 & 5.518  &  2.3042519331774868362  &  0.007347829662205245864     \\

\end{tabular}
\end{center}
\end{table}

\begin{table}[H]
\caption{Bound state and resonances for the potential (\ref{eq:v(x)_WDB2})}
\label{tab:resonances2}
\begin{center}
\par
\begin{tabular}{D{.}{.}{20}D{.}{.}{20}}
\hline  \multicolumn{1}{c}{$\varepsilon_R$ } &
\multicolumn{1}{c}{$-\varepsilon_I$} \\
\hline

 0.5020403621419         &   0    \\
  1.4209709457146932076   &   5.82652808855403\times 10^{-5}        \\
 2.1271970775224959319   &   1.5447312841805183109\times 10^{-2}  \\
 2.5845828598531001914   &   0.17375071916219928095    \\
 2.9244219292377372486   &   0.564794965582576499  \\
 3.255486140023381540    &   1.1115316000246994816    \\
 3.5572161626513698      &   1.7555062346769250   \\
 3.824329026868890       &   2.4874451532278992    \\
 4.055433668209184       &   3.29864420145319      \\
  4.249963938764321       &   4.18316582758871    \\
 4.407748386304          &   5.136439406966     \\
 4.528814027868          &   6.15480966701      \\

\end{tabular}
\end{center}
\end{table}

\begin{thebibliography}{99}
\bibitem{R11}  Rapedius K 2011 \textit{Eur. J. Phys.} \textbf{32} 1199.

\bibitem{DK10}  Dutt A and Kar S 2010 \textit{Am. J. Phys.} \textbf{78} 1352.

\bibitem{WB69}  Wong C Y and Bang J 1969 \textit{Phys. Lett. B} \textbf{29}
143.

\bibitem{CN70}  Cramer J D and Nix J R 1970 \textit{Phys. Rev. C} \textbf{2}
1048.

\bibitem{FT91}  Friedman R S and Truhlar D G 1991 \textit{Chem. Phys. Lett.}
\textbf{183} 539.

\bibitem{RL93}  Ryaboy V and Lefebvre R 1993 \textit{J. Chem. Phys.} \textbf{%
99} 9547.

\bibitem{S39}  Siegert A J F 1939 \textit{Phys. Rev.} \textbf{56} 750.

\bibitem{C68}  Connor J N L 1968 \textit{Mol. Phys.} \textbf{15} 37.

\bibitem{M98}  Merzbacher E 1998 \textit{Quantum Mechanics} (John Wiley \&
Sons, New York).

\bibitem{WHZ82}  Weber T A, Hammer C L, and Zidell V S 1982 \textit{Am. J.
Phys.} \textbf{50} 839.

\bibitem{F95}  Fern\'{a}ndez F M 1995 \textit{J. Phys. A} \textbf{28} 4043.

\bibitem{F96}  Fern\'{a}ndez F M 1996 \textit{J. Phys. A} \textbf{29} 3167.

\bibitem{F11b}  Fern\'{a}ndez F M 2011 \textit{Am. J. Phys.} (in the press)

\bibitem{F11c}  Fern\'{a}ndez F M, Quantum scattering by Wronskians,
arXiv:1101.0957v1 [quant-ph]

\bibitem{F11a}  Fern\'{a}ndez F M 2011 \textit{Eur. J. Phys.} \textbf{32}
723-732.

\bibitem{WC73}  Whitton W N and Connor J N, L. 1973 \textit{Mol. Phys.}
\textbf{26} 1511.

\bibitem{F08}  Fern\'{a}ndez F M, Accurate calculation of eigenvalues and
eigenfunctions. I: Symmetric potentials, arXiv:0807.0655v2 [math-ph]

\bibitem{KM10}  Klaiman S and Moiseyev N 2010 \textit{J. Phys. B} \textbf{43}
185205.

\bibitem{KG02}  Korsch H J and Gl\"uck M 2002 \textit{Eur. J. Phys.} \textbf{%
\ 23} 413.
\end{thebibliography}
\end{document}